\documentclass[aps,prd,superscriptaddress,twoside,showpacs,twocolumn,nofootinbib,10pt,floatfix,showpacs]{revtex4-1}%
\usepackage{verbatim}
\usepackage{bm}
\usepackage{tikz}
\usepackage{multirow}
\usepackage{amsmath}
\usepackage{gensymb}
\usepackage{graphicx}
\usepackage{epsfig}
\usepackage{braket}
\usepackage{subfigure}
\usepackage{ulem}
\usepackage{color}

\allowdisplaybreaks

\begin{document}
\title{Quark model with Hidden Local Symmetry and its application to $T_{cc}$ }
\author{Bing-Ran He}
\email[E-mail: ]{hebingran@njnu.edu.cn}
\affiliation{
Department of Physics, Nanjing Normal University, Nanjing 210023, PR China
}
\author{Masayasu Harada}
\email[E-mail: ]{harada@hken.phys.nagoya-u.ac.jp}
\affiliation{
Department of Physics, Nagoya University, Nagoya, 464-8602, Japan
}
\affiliation{
		Kobayashi-Maskawa Institute for the Origin of Particles and the Universe, Nagoya University, Nagoya, 464-8602, Japan
}
\affiliation{
		Advanced Science Research Center, Japan Atomic Energy Agency, Tokai 319-1195, Japan
}
\author{Bing-Song Zou}
\email[E-mail: ]{zoubs@itp.ac.cn}
\affiliation{
CAS Key Laboratory of Theoretical Physics, Institute of Theoretical Physics, Chinese Academy of Sciences, Beijing 100190, China}
\affiliation{School of Physical Sciences, University of Chinese Academy of Sciences, Beijing 100049, China}
\affiliation{School of Physics, Peking University, Beijing 100871, China}

\date{February 15, 2023}
\begin{abstract}
We propose a chiral quark model including the $\omega$ and $\rho$ meson contributions in addition to the $\pi$ and $\sigma$ meson contributions. 
We show that the masses of the ground state baryons such as the nucleon, $\Lambda_c$ and $\Lambda_b$ are dramatically improved 
in the model with the vector mesons compared with the one without them. 
The study of the tetraquark $T_{cc}$ is also performed in a coupled channel calculation and the resultant mass is much closer to its experimental value than the result without vector meson contribution. 
This approach can be applied to the future study of multi-quark systems.

\end{abstract}
\maketitle

\section{\label{sec:level1}Introduction}
Since the proposal of the quark model,
lots of people spend great efforts to describe meson and baryon in a systematic way (see, e.g., Refs.~\cite{Isgur:1978xj,Bhaduri:1981pn,Isgur:1984bm,Godfrey:1985xj,Capstick:1986ter,Vinodkumar:1999da,Ding:1999xa,Brau:2002zpy}).

\begin{figure}[htp]
	\centering
	\includegraphics[scale=0.52]{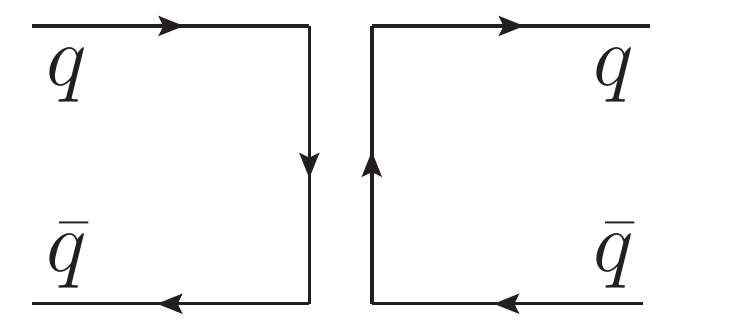}
	\includegraphics[scale=0.52]{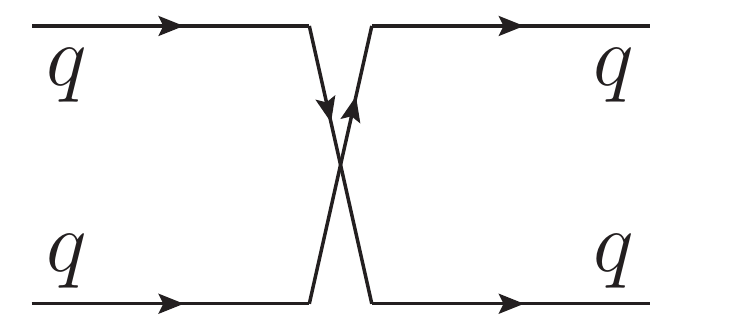}
	\caption{Meson exchange between $q\bar{q}$ (left) and $qq$ (right).
		 }
	\label{fig:feyn_diag}
\end{figure}

Existence of the chiral symmetry and its spontaneous symmetry breaking is one of the most important features in the low-energy region for hadrons including light quarks.
In  Ref.~\cite{Manohar:1983md}, the energy scale of the spontaneous chiral symmetry breaking (S$\chi$SB) is shown to be larger than the confinement scale, and
a chiral quark model was proposed in which the chiral symmetry is spontaneously broken to generate the quark mass and the pion as the Nambu-Goldstone boson associated with the S$\chi$SB. The model includes the pion in addition to the gluon which provides the color force including the confining effect.
In the chiral quark model, the pseudoscalar mesons can be exchanged between a quark and an anti-quark as well as between two quarks, which is understood as shown in Fig.~\ref{fig:feyn_diag}.
Since the exchanged mesons are classified by the chiral symmetry, such models are generally called chiral quark models.

There are several scenarios to construct a chiral quark model. 
Shimizu~\cite{Shimizu:1984iel} studied the quark exchange effects by using pseudo-scalar exchange and confining potential(CON). 
Obukhovsky and  Kusainov~\cite{Obukhovsky:1990tx}
employed scalar and pseudo-scalar exchange, one gluon exchange(OGE) and CON to study nucleon nucleon scattering and the baryon spectrum. 
Glozman and Riska~\cite{Glozman:1995fu,Glozman:1999vd} included pseudo-scalar and vector meson exchange and also CON to study baryon spectrum.  
Dai \textit{et al.}~\cite{Dai:2003dz} included scalar, pseudo-scalar and vector meson exchange together with OGE and CON to study the phase shifts of NN scattering. They found that when vector meson is included, the phase shift is obviously improved in $^1S_0$ channel.  
Vijande \textit{et al.}~\cite{Vijande:2004he,Vijande:2009pu} and Valcarce \textit{et al.}~\cite{Valcarce:2005em} included scalar, pseudo-scalar, OGE and CON to study meson and baryon spectrum. They did not include the vector mesons for avoiding the double counting. 
It was stressed~\cite{Vijande:2004he,Vijande:2009pu} that the chiral quark model beautifully reproduces the mass spectra of mesons constructed from light and heavy quarks, and also the nucleon-nucleon scattering phase shift. 
However, it is known that spectra of baryons are hard to be reproduced since the ``$\sigma$ meson" provides too much strong attractive force between two quarks~\cite{Valcarce:2005em}.
Actually, if the best fitted parameters of the chiral quark model shown in Ref.~\cite{Vijande:2004he} were used, three ground states of baryons would get extremely low values than experimental results, i.e., $M_N^{\rm(exp.)}-M_N^{\rm(theo.)}=262$\,MeV,  $M_{\Lambda_c}^{\rm(exp.)}-M_{\Lambda_c}^{\rm(theo.)}=322$\,MeV and $M_{\Lambda_b}^{\rm(exp.)}-M_{\Lambda_b}^{\rm(theo.)}=359$ MeV.

After the observation of $X(3872)$ in 2003, lots of exotic states are observed experimentally and studied theoretically (see, e.g., Refs.~\cite{Brambilla:2010cs,Esposito:2014rxa,Chen:2016spr,Esposito:2016noz,Chen:2016qju,Guo:2017jvc,Olsen:2017bmm,Liu:2019zoy,Brambilla:2019esw,Chen:2022asf}).
Recently the discovered $T_{cc}$ state~\cite{LHCb:2021vvq,LHCb:2021auc} provides a new opportunity to check the validity of the chiral quark model.
Actually, a problem similar to the one for the above baryons happens in the study of the tetraquark $T_{cc}$: the diquark picture of $T_{cc}$ gives very deep binding energy and the resultant mass is far below the $D D^\ast$ threshold energy, $\delta m = M_{T_{cc}} - M_D - M_{D^\ast} \sim - (130\mbox{-}185)\,\mbox{MeV}$~\cite{Pepin:1996id,Vijande:2003ki,Yang:2009zzp,Deng:2018kly,Yang:2019itm,Tan:2020ldi} to be compared wth $-0.36\pm 0.04^{+0.004}_{-0}$\,MeV of the experimental observation~\cite{LHCb:2021auc}.
We note that the wave function of
the $T_{cc}$ and the above three baryons
contain a ``good diquark'' constructed from two light quarks.

The problem existing in $I=0$ and $S=0$ light diquark channel indicates some important interactions are missing in the chiral quark model.
We notice that, in Ref.~\cite{Manohar:1983md}, the chiral symmetry breaking scale is estimated as about $\Lambda_\chi \sim 1.2\,$GeV, and that the chiral quark model is constructed as an effective field theory applicable below $\Lambda_\chi$. 
The inclusion of vector meson  is studied in lots of effective models, e.g., one boson exchange model in nucleon interaction~\cite{Moravcsik_1972,Meissner:1987ge,Machleidt:2017vls}, 
Skyrme model~\cite{Broniowski:1984zd,Fujiwara:1984pk,Igarashi:1985et,Broniowski:1985kj,Meissner:1986vu,Birse:1986qc,Meissner:1987ge} in addition to the quark models mentioned above~\cite{Glozman:1995fu,Glozman:1999vd,Dai:2003dz}. 
Furthermore, in the framework of the hidden local symmetry (HLS)~\cite{Meissner:1987ge,Bando:1987br,Harada:2003jx}, these vector mesons are strongly related to the spontaneous chiral symmetry breaking. 

In this paper, we  
construct an effective field theory for the chiral quark including the vector mesons in addition to scalar and pseudoscalar mesons using the framework of the HLS, to study mesons, baryons and also exotic states.
We note that, 
in the HLS, the model possesses the chiral symmetry without including the axial-vector mesons~\cite{Bando:1987br,Harada:2003jx}. 
We will show that 
the $\omega$ meson provides the attractive force between a quark and an anti-quark,
while  repulsive force is generated between two quarks due to its $G$-parity.
This property is contrasted to the one for the colored force: the colored force gives attractive force in both channels (although strengths are different).
Thus, we expect that the inclusion of the $\omega$ meson with changing the colored force
cures the light diquark problem in the chiral quark model, while the success for the light mesons is kept.
We should note that, since the $\omega$ meson is not exchanged by mesons including a heavy quark, we only modify the the colored force in the middle and long range.
Furthermore, since the gluon and vector mesons play different roles in $q \bar{q}$ and $qq$ channels, we expect that the double counting is avoided by adjusting coupling constants of the gluon exchange and vector meson exchange.

\section{\label{sec:level2}The Quark Model with Hidden Local Symmetry}
In this work, we only discuss the $u,d,c,b$ quarks for simplicity.
We leave the inclusion of the strange quark in future work~\cite{He:2022su3}.
Now, the Hamiltonian of the present model is written as
\begin{eqnarray}\label{eq:H}
	H&=&\sum_{i=1}\left(m_i+\frac{p_i^2}{2m_i}\right)-T_{CM} + \sum_{j>i=1}\left(V^{\rm CON}_{ij}\right.\nonumber\\
	&&\left.+V^{\rm OGE}_{ij}+V_{ij}^{\sigma}+V_{ij}^{\pi}+V_{ij}^{\omega}+V_{ij}^{\rho}\right)\,,
\end{eqnarray}
where $m_{i}$ and $p_{i}$ are the mass and the momentum of $i$-{th} quark, $T_{CM}$ is the kinetic energy of the center of mass of the system. $V^{\rm CON}_{ij}$, $V^{\rm OGE}_{ij}$, $V_{ij}^{\sigma}$ and $V_{ij}^{\pi}$, represent the potential of confinement, one-gluon-exchange, $\sigma$ and $\pi$ exchange, which are given in Refs.~\cite{Vijande:2004he,Valcarce:2005em}.
$V_{ij}^{\omega}$ and $V_{ij}^{\rho}$ are the potential of $\omega$ and $\rho$ exchange.
They take the same form except the iso-spin dependence. It is convenient to define the common part $V_{ij}^{v=\omega,\rho}$ as
$V_{ij}^{\omega}=V_{ij}^{v=\omega}$ and $V_{ij}^{\rho}={\boldsymbol\tau}_i \cdot{\boldsymbol\tau}_j V_{ij}^{v=\rho}$ for $qq$ ($q=u,d$) potential, where
$\boldsymbol \tau_i$ is the flavor $SU(2)$ Pauli matrices of $ i$-{th} quark.
The $V_{ij}^{v}$ in the spatial coordinate is given as
\begin{eqnarray}	
	V_{ij}^{v}&=&\frac{\Lambda_v^2}{\Lambda_v^2-m_v^2}\left\{ \frac{g_v^2}{4\pi}
	m_v\left[Y(m_v r)-\left(\frac{\Lambda_v}{m_v}\right)Y(\Lambda_v r)\right]\right.\nonumber\\
	&&+\frac{m_v^3}{ m_i m_j}\left(\frac{g_v(2f_v+g_v)}{16\pi}+\frac{{\boldsymbol \sigma}_i\cdot {\boldsymbol \sigma}_j}{6}\frac{(f_v+g_v)^2}{4\pi} \right)\nonumber\\
	&&\;\;\times\left[ Y(m_v r)-\left(\frac{\Lambda_v}{m_v}\right)^3 Y(\Lambda_v r)\right] \nonumber\\
	&&-{\boldsymbol S}_{+}\cdot{\boldsymbol L} \frac{g_v(4f_v+3g_v)}{8\pi}\frac{m_v^3}{ m_im_j} \nonumber\\
	&&\;\;\times\left[G(m_v r)-\left(\frac{\Lambda_v}{m_v}\right)^3 G(\Lambda_v r)\right] \nonumber\\
	&& - {\boldsymbol S}_{ij} \frac{(f_v+g_v)^2}{4\pi} \frac{m_v^3}{12m_i m_j}\nonumber\\
	&&\;\;\left.\times\left[H(m_v r)-\left(\frac{\Lambda_v}{m_v}\right)^3 H(\Lambda_v r)\right] \right\}\,.
	\label{eq:v_vec}
\end{eqnarray}
Here $m_v$, $\Lambda_v$, $g_v$ and $f_v$ are the mass, cutoff, electric and magnetic coupling constants
of the relevant vector meson, respectively. $\boldsymbol \sigma_i$ is the spin $SU(2)$ Pauli matrices of $ i$-{th} quark,
${\boldsymbol S}_{+}=\frac{{\boldsymbol \sigma}_i+{\boldsymbol \sigma}_j}{2}$, ${\boldsymbol S}_{ij}=3({\boldsymbol \sigma}_i\cdot\hat{r}_{ij})({\boldsymbol \sigma}_j\cdot\hat{r}_{ij})-{\boldsymbol \sigma}_i\cdot{\boldsymbol \sigma}_j$.
$Y(x)=e^{-x}/x$, $H(x)=(1+3/x+3/x^2)Y(x)$ and $G(x)=(1/x+1/x^2)Y(x)$.
The potentials of $V_\omega$ and $V_\rho$ for $q\bar{q}$ ($q=u,d$) is obtained by performing a $G$-parity transformation of that in $q{q}$ case.  

We solve the Schr\"odinger equation of the Hamiltonian (\ref{eq:H}) by using Gaussian expansion method (GEM)~\cite{Hiyama:2003cu}. 
For studying the effects of vector mesons, we keep the model parameters of the $\pi$ and $\sigma$ potentials as Ref.~\cite{Vijande:2004he}.
We fit the model parameters of vector meson potentials and color potential to several mesons and baryons.
We show best fitted values of model parameters in Table~\ref{tab:para}.
\begin{table}[htp]
	\caption{Model Parameters.
		The naming scheme is the same as Ref.~\cite{Vijande:2004he}.}\label{tab:para}
	\begin{ruledtabular}
		\begin{tabular}{ c c  c c}
			$m_u = m_{d}({\rm MeV})$ & 303.3 & $m_{\sigma}({\rm fm}^{-1})$ & 3.42 \\
			$m_c({\rm MeV})$ & 1696.0 &  $m_{\pi}({\rm fm}^{-1})$ & 0.7 \\
			$m_b({\rm MeV})$ & 5039.6 & $m_{\omega}({\rm fm}^{-1})$ & 3.97 \\
			$a_c({\rm MeV})$ & 364.5 & $m_{\rho}({\rm fm}^{-1})$ & 3.93 \\
			$\mu_c({\rm fm}^{-1})$ & 0.6 & $\Lambda_{\sigma} = \Lambda_{\pi}({\rm fm}^{-1})$ & 4.2 \\
			$\Delta({\rm MeV})$ & 100.28
			& $\Lambda_{\omega}=\Lambda_{\rho}({\rm fm}^{-1})$ & 7.2 \\
			$a_s$ & 0.731
			 & $g_{ch}^2/(4\pi)$  & 0.54 \\
			$\alpha_0$ & 2.742 & $g_{\omega}$ & 3.841 \\
			$\Lambda_0({\rm fm}^{-1})$ & 0.304 & $g_{\rho}$ & 0.675 \\
			$\mu_0({\rm MeV})$ & 241.205 &  $f_{\omega}$ & -1.416 \\
			$\hat{r}_0({\rm MeV} \cdot {\rm fm})$ & 95.621 & $f_{\rho}$ & 0.581 \\
			$\hat{r}_g({\rm MeV} \cdot {\rm fm})$ & 155.066 &  & 
			\\
		\end{tabular}
	\end{ruledtabular}
\end{table}
Here, we newly introduce eight parameters, i.e., $m_\omega$, $m_\rho$, $\Lambda_\omega$, $\Lambda_\rho$, $g_\omega$, $g_\rho$, $f_\omega$ and $f_\rho$,
with other parameters in the
same naming scheme as Ref.~\cite{Vijande:2004he}.
The vector mass $m_{\omega}$ and $m_{\rho}$ are taken from PDG~\cite{ParticleDataGroup:2022pth}.
The cut off $\Lambda_{\omega}$ and $\Lambda_{\rho}$ of vector meson are not sensitive to get the result, here we take a typical value of $7.2 \,{\rm fm^{-1}}$. 
The wave functions for meson and baryon can be found in~\cite{Amsler:2018zkm}.

\section{\label{sec:level3}Results}
\begin{figure}[htp]
	\centering
	\includegraphics[scale=0.32]{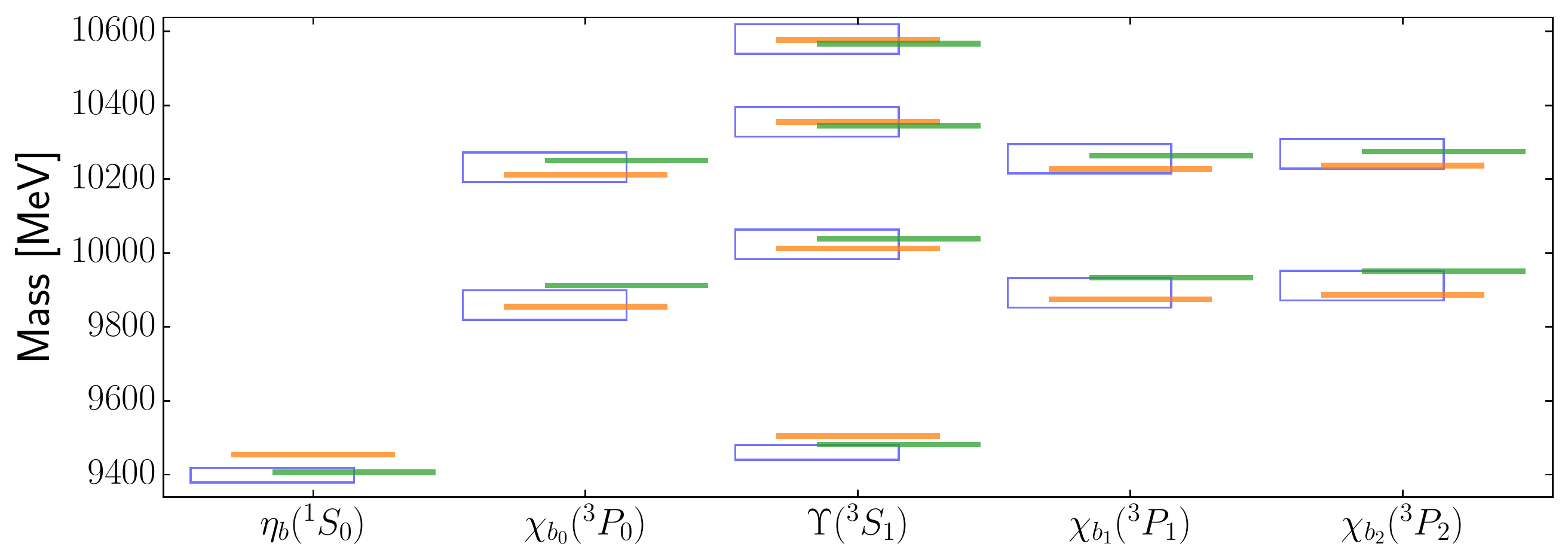}
	\includegraphics[scale=0.32]{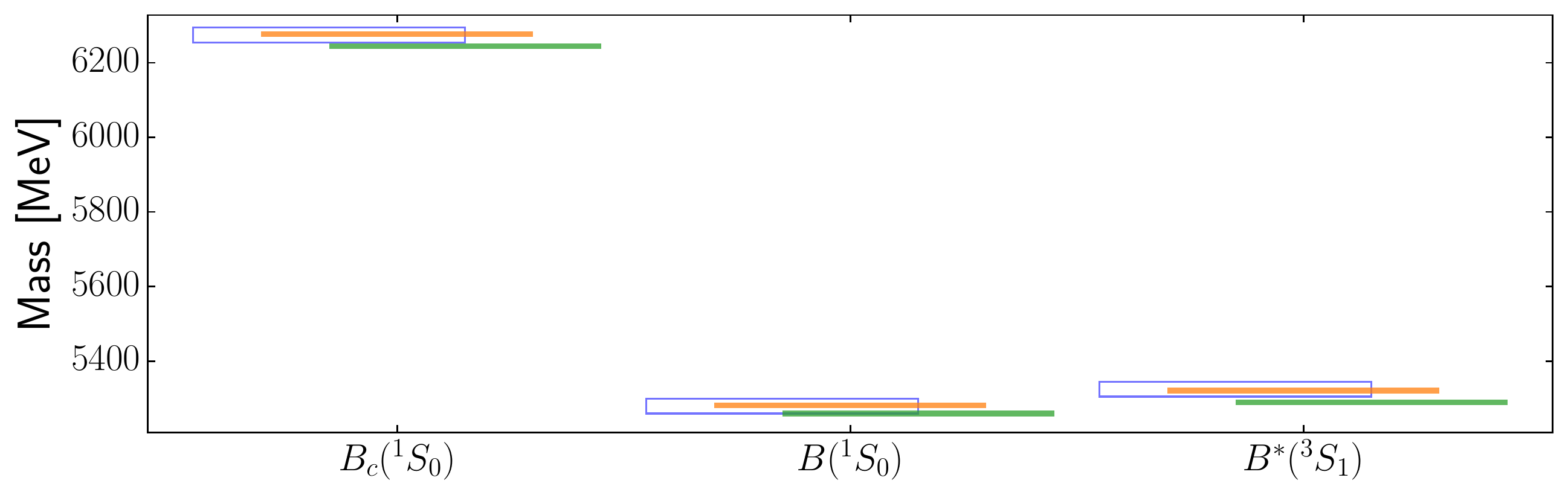}
	\includegraphics[scale=0.32]{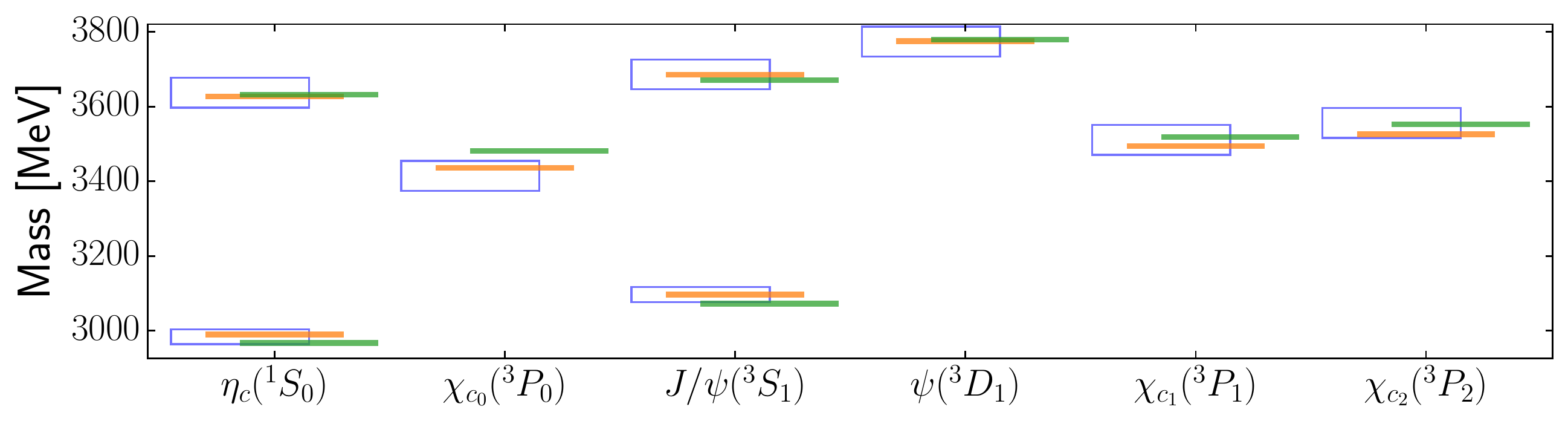}
	\includegraphics[scale=0.32]{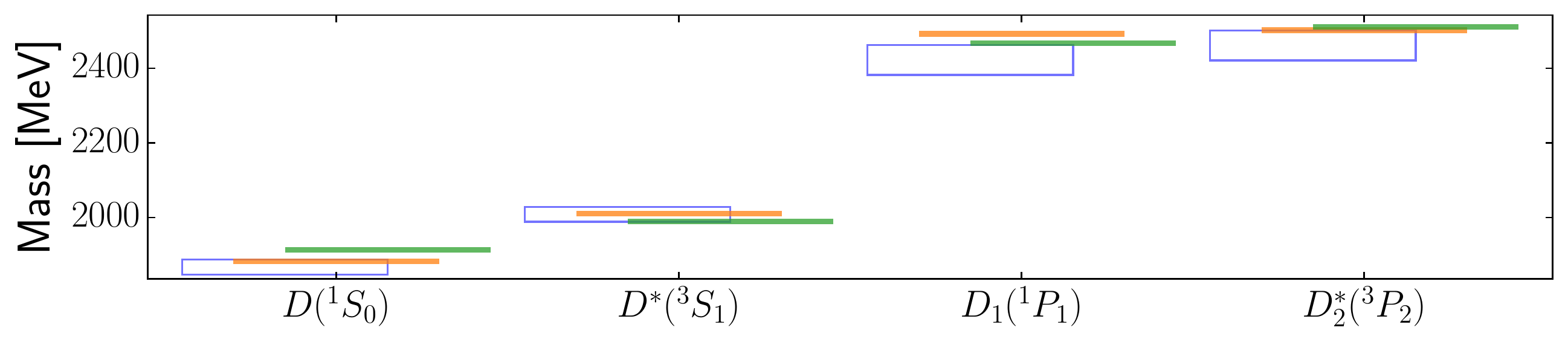}
	\includegraphics[scale=0.32]{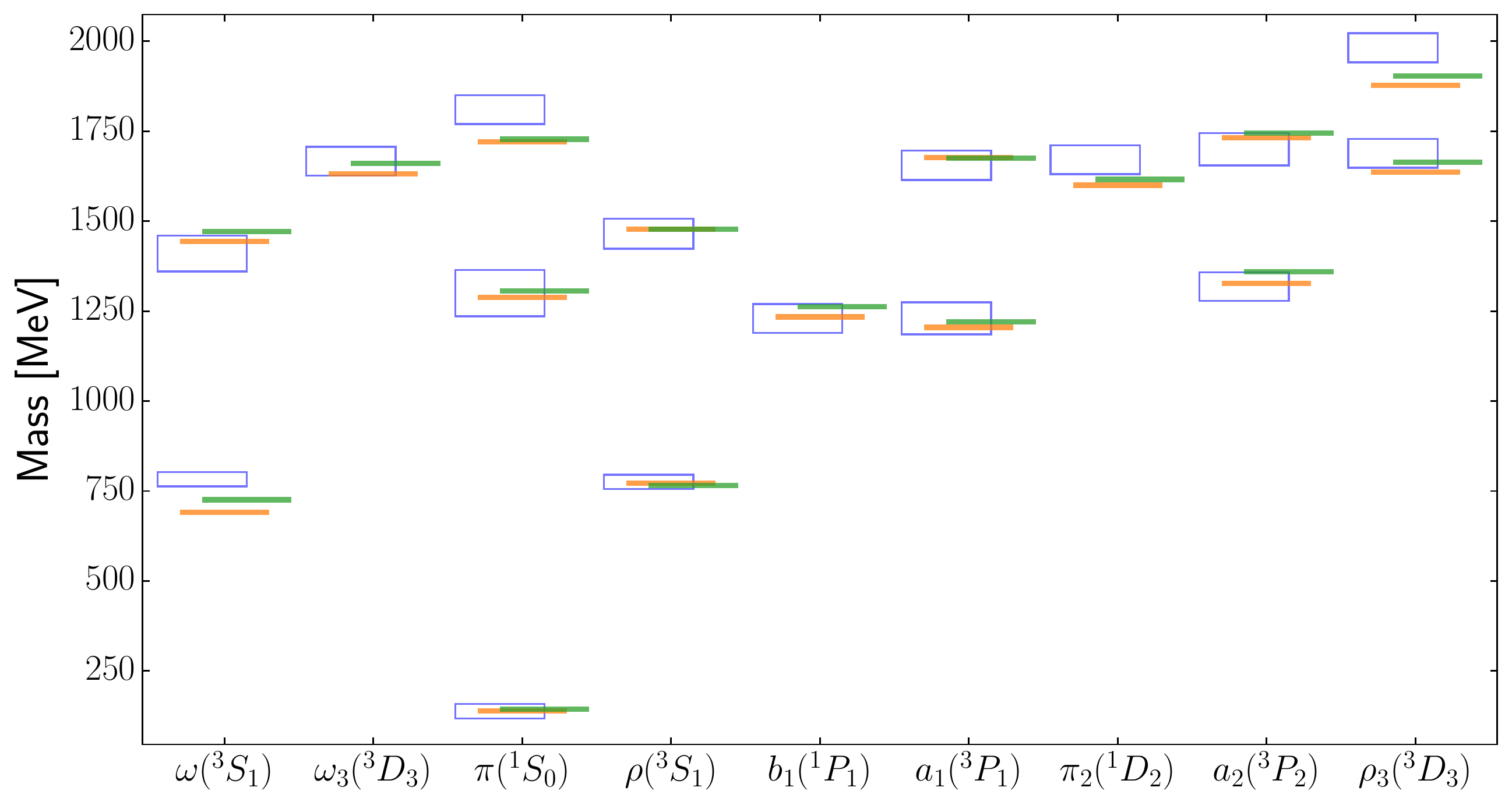}
	\caption{Mass spectrum in $b\bar{b}$ (1st panel),
	$c\bar{b}$ and $b\bar{q}$ (2nd panel), $c\bar{c}$ (3rd panel), $q\bar{c}$ (4th panel) and $q\bar{q}$ (5th panel) systems. The green lines show the predictions in the present model including vector mesons, while the orange lines the ones in the model without vector mesons.}
	\label{fig:meson_mass_1}
\end{figure}

\begin{figure}[htp]
	\centering
	\includegraphics[scale=0.32]{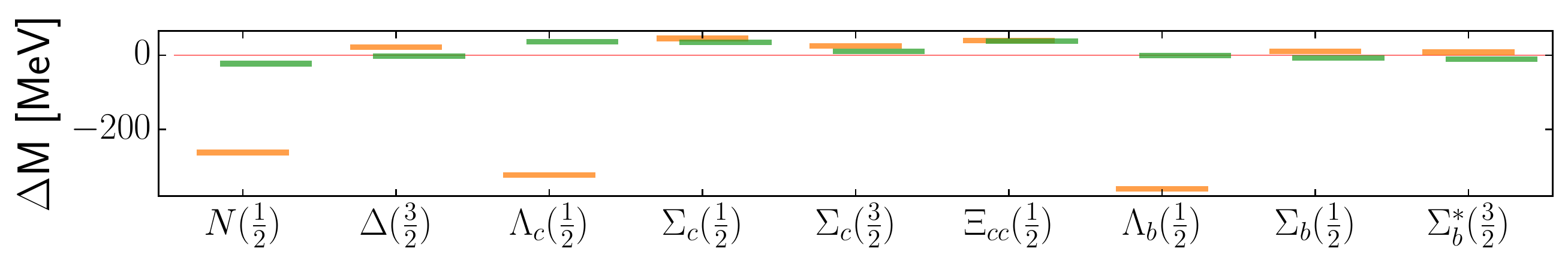}
	\caption{Mass spectrum of baryon system. The meaning of colors are as in Fig.~\ref{fig:meson_mass_1}.}
	\label{fig:baryon_mass_1}
\end{figure}

We show the meson masses calculated in the present model in 
Fig.~\ref{fig:meson_mass_1}, and baryon masses in Fig.~\ref{fig:baryon_mass_1}.

The blue block represents the error calculated as
${\rm Err(sys)}=\sqrt{{\rm Err(exp)}^2+{\rm Err(the)}^2}$, where
${\rm Err(exp)}$ is the experimental error taken from PDG~\cite{ParticleDataGroup:2022pth}, while ${\rm Err(the)}$ represents
the model limitation error as we do not include isospin breaking effects, and also ignore mixing effects (e.g. S-D, P-F mixing between meson states).
We take ${\rm Err(the)}$ as about $40$\,MeV for ground-state and $80$\,MeV for excited-state of mesons and baryons. We minimize $\chi^2=\sum_{i}(\frac{m_i({\rm the})-m_i({\rm exp})}{{\rm Err_i(sys)}})^2$ of the system, where $m_i({\rm the})$ and $m_i({\rm exp})$ are theoretical and experimental mass of each particle, respectively. 
The total $\chi^2$ of the present work is about 15.12. 
The degree of freedom (dof) of present work is $\mbox{dof}=51-(23-3)=31$, which is calculated in the following way: (i) we used 51 particles (42 mesons and 9 baryons) to minimize $\chi^2$, (ii) the total parameter of present model is 23 but we fixed 3 parameters $m_\pi$, $m_\omega$ and $m_\rho$ as experiment values.  
Thus $\chi^2/\mbox{dof} \simeq 15.12/31 \simeq 0.4877$ indicate the reliability of the present model is about $99.26\%$. 
The orange blocks for mesons are taken from Ref.~\cite{Vijande:2004he}, the corresponding baryon spectrum are calculated using the same parameter. The green blocks show the predictions of mass spectra of the present model.

We would like to stress that the spectra of ground-state baryons $N$, $\Lambda_c$ and $\Lambda_b$,  which all contain a good diquark,
are dramatically improved as one can easily see from Fig.~\ref{fig:baryon_mass_1}, while the present model reproduces the meson spectra as good as the original model in Ref.~\cite{Vijande:2004he} as shown in Fig.~\ref{fig:meson_mass_1}.
From the values of parameters listed in Table~\ref{tab:para}, this is the consequence of the $\omega$ exchange contribution:
As we stated above, the $\omega$ meson provides the attractive force between a quark and an anti-quark and the repulsive force between two quarks.

Here, to understand Table~\ref{tab:para} we clarify the operators included in the potentials with the sign of each contribution, which are summarized in Table~\ref{tab:oper}.
\begin{table}[htp]
\caption{\label{tab:oper}List of operators included in the potentials for $L=0$ states and the sign of the contributions.
The left (right) side of slash represents $qq$ ($q\bar{q}$), respectively.
}
	\begin{ruledtabular}
		\begin{tabular}{ c c  c c c}
 & 1     & $\tau_i\tau_j$ & $\sigma_i\sigma_j$ & $\sigma_i\sigma_j\cdot\tau_i\tau_j$ \\
 $\sigma$~\cite{Vijande:2004he} & $-/-$ &                &                    &                                     \\
$\pi$~\cite{Vijande:2004he}    &       &                &                    & $+/-$                               \\
$a_0$~\cite{Vijande:2009pu}    &       & $-/+$          &                    &                                     \\
OGE~\cite{Vijande:2004he}      & $-/-$ &                & $+/+$              &                                     \\
CON~\cite{Vijande:2004he}      & $+/+$ &                &                    &                                     \\
$\omega$ (This work)           & $+/-$ &                & $+/-$              &                                     \\
$\rho$ (This work)             &       & $+/+$          &                    & $+/+$
		\end{tabular}
	\end{ruledtabular}
\end{table}
Here we only discuss $L=0$ states for simplicity, in such case only central force remains.
The problem of deep binding energy occurs when good light diquark appears ($I=0$ and $S=0$).
To solve this problem, we have four possible ways of modification to provide a repulsive force between $qq$: (i) modify terms proportional to the operator `$1$' in all $I$ and $S$ channel, (ii) modify terms to $\tau_i\tau_j$ in $I=0$ channel, (iii) modify terms to $\sigma_i\sigma_j$ in $S=0$ channel, (iv) modify terms to $\sigma_i\sigma_j\cdot\tau_i\tau_j$ in $I=0$ and $S=0$ channel.
We should note that we kept $V_\pi$ and $V_\sigma$ as same as in Ref.~\cite{Vijande:2004he}, since
Ref.~\cite{Vijande:2004he} has already given good description of meson spectra and also phase shift of baryon-baryon scattering.
This implies that $g_\rho + f_\rho$ in $V_\rho$ cannot be too big since it generates the contribution proportional to $\sigma_i\sigma_j\cdot\tau_i\tau_j$ operator which is already saturated by the contribution from $V_\pi$, and thus the way (iv) is ruled out.
The modification of the way (ii) is done by $V_\rho$ (terms proportional to $g_\rho$) playing a role similar to the one by $a_0$ meson exchange, which is introduced to explain the $\omega$-$\rho$ masses in Ref.~\cite{Vijande:2009pu}.
Since $a_0$ only give small modification of meson spectra, as can be seen from
the predicted mass $\omega(^3S_1)$ in Fig.~\ref{fig:meson_mass_1}, $g_\rho$ should be very small.
As a result there remains only two ways, i.e., (i) and (iii), which is achieved only by the inclusion of $\omega$ meson contribution proportional to $g_\omega$ and $f_\omega + g_\omega$.
Here, let us briefly explain why the $\omega$ meson contribution can cure the problem of deep binding of good light diquark in $S=0$ channel.
As seen in Table~\ref{tab:oper}, the one-gluon exchange potential $V^{\rm OGE}$ contains the $\sigma_i\sigma_j$ operator, which gives the attractive force in both $q\bar{q}$ and $qq$ channels, while the $\sigma_i\sigma_j$ term in $V_\omega$ generates attractive force in $q\bar{q}$ channel but repulsive force in $qq$ channel.
For the operator `$1$', the change of $V^{\rm OGE}$ and $V_\omega$ can be compensated by $V^{\rm CON}$, i.e., inclusion of $\omega$ together with the weakening of color potential make the attractive force in $qq$ channel smaller with keeping the force in $q\bar{q}$ channel intact.
As a conclusion, the way (i) and (iii) with taking $\vert g_\rho \vert \sim \vert f_\rho \vert  \sim 0$ is the minimal modification to the model in Ref.~\cite{Vijande:2004he}.

\begin{figure}[htp]
	\centering
	\includegraphics[scale=0.15]{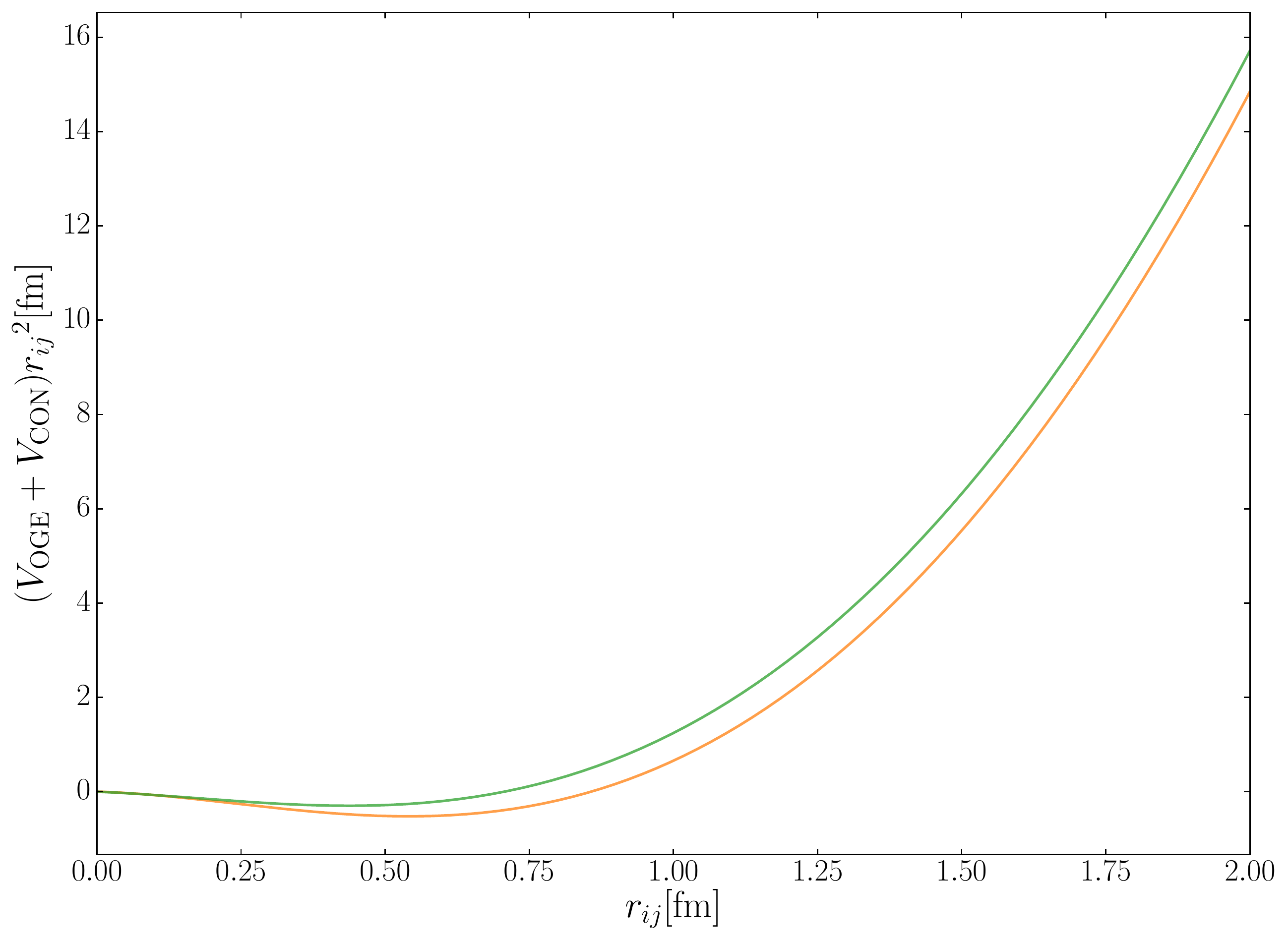}
	\includegraphics[scale=0.15]{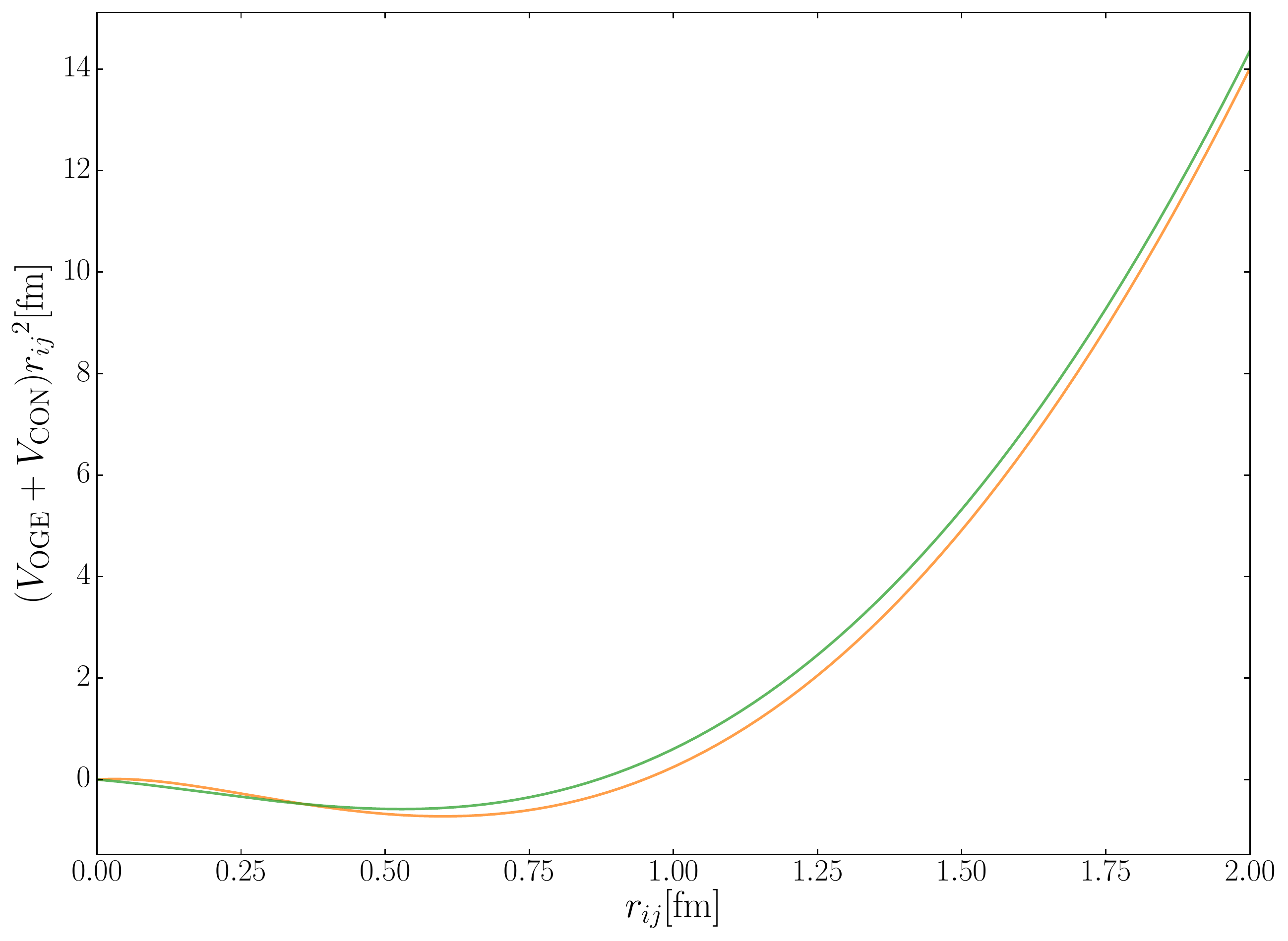}
	\caption{Potential $(V_{\rm OGE}+V_{\rm CON})r^2$ of $\Upsilon$(left) and $\omega$(right). Orange and green lines represent  Ref.~\cite{Vijande:2004he} and present study, respectively.}
	\label{fig:potential_upsilon}
\end{figure}

We next explain a mechanism why the spectra of mesons including a heavy quark,
$c\bar{c}$, $q\bar{c}$ and $b\bar{b}$ systems, are as good as in Refs.~\cite{Vijande:2004he,Valcarce:2005em} although the color potential $V^{\rm OGE} + V^{\rm CON}$, which is the only force acting in these systems, is modified.
Since the effective range of heavy quark sector is rather smaller than that of light sector, we can modify the color potential $V^{\rm CON}_{ij} +V^{\rm OGE}_{ij}$  in middle and long range part without changing the short range part to maintain the main features of mesons made of heavy quarks.  In the present work we achieved this modification by keeping the forms  in Ref.~\cite{Vijande:2004he} and just changing the values of parameters.
In Fig.~\ref{fig:potential_upsilon}, we show the color potential $(V_{\rm OGE}+V_{\rm CON})r^2$ of $\Upsilon$(left) and $\omega$(right).
This shows that the color potential is changed slightly in both heavy and light quark sectors when the distance is larger than about $0.25$ fm.
Typical distance between two heavy quarks such as $\Upsilon(1S)$ in the present analysis is about $0.21$\,fm, so that the potential relevant for mesons made from $b$ and $\bar{b}$ is not changed.
Thus, we can see that, from Fig.~\ref{fig:meson_mass_1}
the spectra of mesons including
heavy quarks are in reasonable agreement with experiment.
On the other hand, a typical distance between $q$ and $\bar{q}$ in mesons of the ground state, e.g., $\rho(1S)$ is about $0.83$\,fm, so that the color attractive force is actually weakened.
The attractive force mediated by the $\omega$ meson between $q$ and $\bar{q}$ compensates the weakening effect of color force to provide good agreement of the predictions of the present model with experiment as can be seen in Fig.~\ref{fig:meson_mass_1}.
We note that typical distances between $c\bar{c}$ system and $q\bar{c}$ system are about $0.37$\,fm ($J/\psi$) and  $0.64$\,fm ($D^*$).
Then, the spectra of $c\bar{c}$ and $q\bar{c}$ systems are not as good as the one of $b\bar{b}$ system.
For further improvement
of the spectrum we might need to modify the forms of color potential, which we leave as a future work.

Based on the above success in the meson and baryon sectors, we now perform a coupled channel  calculation of $T_{cc}$ and $T_{bb}$ tetraquarks. 
The $T_{cc}$ and $T_{bb}$ have eight channels, with six of meson-meson structure and two of diquark--anti-diquark structure. The detailed wave functions of $T_{cc}$ and $T_{bb}$ can be found in Refs.~\cite{Yang:2019itm,Tan:2020ldi}.

\begin{table}[htp]
	\caption{\label{tab:tcc_tbb}Mass spectrum of $T_{cc}$ and $T_{bb}$. 
		The subscript of channel denotes the color wave functions, $1 \otimes 1$, $8 \otimes 8$, $\bar{3} \otimes 3$ and $6 \otimes \bar{6}$ denote color singlet-singlet, octet-octet, antitriplet-triplet and sextet-antisextet, respectively. 
		The binding energy(MeV) is $E_B=M-E_{th}$, where $M$ is the tetra-quark mass and $E_{th}$ is the threshold of $T_{cc}$ and $T_{bb}$, i.e., $E_{th}(DD^*)=3903.1\mbox{\,MeV}$ and $E_{th}(BB^*)=10549.4\mbox{\,MeV}$. 
		The 3-rd and 6-th column are the component percentage in channel coupling calculation. 
	}
	\begin{ruledtabular}
		\begin{tabular}{ c  r c c r c c}
			Channel                                          & $E_B$ &        & Channel                                          & $E_B$ &        &  \\ \hline
			$[DD^*]{}_{1 \otimes 1}$                         & $6.2$  & $39\%$ & $[BB^*]{}_{1 \otimes 1}$                         & $0.5$   & $12\%$ &  \\
			$[D^*D]{}_{1 \otimes 1}$                         & $6.2$      &  $39\%$      & $[B^*B]{}_{1 \otimes 1}$                         & $0.5$      & $12\%$       &  \\
			$[D^*D^*]{}_{1 \otimes 1}$                       &  $83.1$     & $5\%$       & $[B^*B^*]{}_{1 \otimes 1}$                       &  $31$     & $32\%$       &  \\
			$[DD^*]{}_{8 \otimes 8}$                         & $383.1$ &  $0\%$      & $[BB^*]{}_{8 \otimes 8}$                         & $253.6$   & $0\%$       &  \\
			$[D^*D]{}_{8 \otimes 8}$                         &  $383.1$     & $0\%$       & $[B^*B]{}_{8 \otimes 8}$                         & $253.6$      & $0\%$       &  \\
			$[D^*D^*]{}_{8 \otimes 8}$                       & $337.3$      & $0\%$       & $[B^*B^*]{}_{8 \otimes 8}$                       & $233.3$      & $0\%$       &  \\
			$[(cc)(\bar{q}\bar{q})^*]{}_{6 \otimes \bar{6}}$ & $337.5$      & $0\%$       & $[(bb)(\bar{q}\bar{q})^*]{}_{6 \otimes \bar{6}}$ & $233.7$      & $0\%$       &  \\
			$[(cc)^*(\bar{q}\bar{q})]{}_{\bar{3} \otimes {3}}$ &  $120.3$     &  $17\%$      & $[(bb)^*(\bar{q}\bar{q})]{}_{\bar{3} \otimes {3}}$ &   $-37.8$    & $44\%$       &  \\
			\hline
			Mixed                                            & $-4.9$     &        & Mixed                                            &   $-88.2$    &        &
		\end{tabular}
	\end{ruledtabular}
\end{table}

\begin{table}[htp]
	\caption{\label{tab:rms_tcc_tbb}RMS distance (fm) between two quarks of $T_{cc}$ and $T_{bb}$ in coupled channel calculation. }
	\begin{ruledtabular}
		\begin{tabular}{ c c  c c c c }
			& $r_{cc}$ & $r_{\bar{q}c}$ & $r_{\bar{q}\bar{q}}$ & $r_{\bar{q}b}$ & $r_{bb}$ \\ \hline
			$T_{cc}$ & $1.56$         &  $1.24$              &  $1.70$                    &                &          \\
			$T_{bb}$ &          &                &  $0.75$                    &  $0.65$              & $0.37$
		\end{tabular}
	\end{ruledtabular}
\end{table}

We summarize the results of $T_{cc}$ and $T_{bb}$ in coupled channel calculation in Table~\ref{tab:tcc_tbb}. 
We find that $T_{cc}$ cannot form a bound state in single channel calculation, while after channel coupling, the mass of $T_{cc}$ is $4.9$\,MeV below the $DD^\ast$ threshold, which is much improved compared with the previous result of $180$\,MeV. 
For $T_{bb}$ system, diquark--anti-diquark structure of $\bar{3} \otimes {3}$ will give a bind state of $37.8$ MeV below the $BB^\ast$ threshold, and after channel coupling, the mass of $T_{bb}$ is predicted to be $88.2$\,MeV below the $BB^\ast$ threshold.  
This is understandable because the distance between $bb$ is smaller than that between $cc$ and feel stronger color electric force which is proportional to $1/r$. 
The root mean square (RMS) distance between two quarks included in $T_{cc}$ and $T_{bb}$ are given in Table~\ref{tab:rms_tcc_tbb}. 
We conclude that $T_{cc}$ is like a meson-meson molecular state, while $T_{bb}$ is a compact tetraquark state.

\section{\label{sec:level4}summary}
We add the vector meson exchange effects to a chiral quark model to produce a repulsive fore between two light quarks.
This effect dramatically improve the spectra of the baryons including a good diquark, while meson spectra are in good agreement with experiments.
In addition, the predicted mass of $T_{cc}$ is much closer to its experimental value than the result without vector mesons. 
This approach can be applied to the future study of multi-quark systems.

\begin{acknowledgments}
B.R.~He was supported in part by the National Natural Science Foundation of China (Grant Nos. 11705094 and 12047503), Natural Science Foundation of Jiangsu Province, China (Grant No. BK20171027), Natural Science Foundation of the Higher Education Institutions of Jiangsu Province, China (Grant Nos. 17KJB140011 and 22KJB140012).

B.S. Zou was supported by the National Natural Science Foundation
of China (NSFC) and the Deutsche Forschungsgemeinschaft
(DFG, German Research Foundation) through the funds provided to
the Sino-German Collaborative Research Center TRR110 Symmetries
and the Emergence of Structure in QCD (NSFC Grant No.
12070131001, and DFG Project-ID 196253076-TRR 110), the NSFC
(11835015, and 12047503), and the Grant of Chinese Academy of
Sciences (XDB34030000).

M.~Harada was supported in part by JSPS KAKENHI Grant No. 20K03927.
\end{acknowledgments}




\begin{thebibliography}{99}

\bibitem{Isgur:1978xj}
N.~Isgur and G.~Karl,
Phys. Rev. D \textbf{18}, 4187 (1978).

\bibitem{Bhaduri:1981pn}
R.~K.~Bhaduri, L.~E.~Cohler and Y.~Nogami,
Nuovo Cim. A \textbf{65}, 376-390 (1981).

\bibitem{Isgur:1984bm}
N.~Isgur and J.~E.~Paton,
Phys. Rev. D \textbf{31}, 2910 (1985).

\bibitem{Godfrey:1985xj}
S.~Godfrey and N.~Isgur,
Phys. Rev. D \textbf{32}, 189-231 (1985).



\bibitem{Capstick:1986ter}
S.~Capstick and N.~Isgur,
Phys. Rev. D \textbf{34}, no.9, 2809-2835 (1986).

\bibitem{Vinodkumar:1999da}
P.~C.~Vinodkumar, J.~N.~Pandya, V.~M.~Bannur and S.~B.~Khadkikar,
Eur. Phys. J. A \textbf{4}, 83-90 (1999).

\bibitem{Ding:1999xa}
Y.~B.~Ding, X.~Q.~Li and P.~N.~Shen,
Eur. Phys. J. A \textbf{7}, 107-108 (2000).

\bibitem{Brau:2002zpy}
F.~Brau, C.~Semay and B.~Silvestre-Brac,
Phys. Rev. C \textbf{66}, 055202 (2002).

\bibitem{Manohar:1983md}
A.~Manohar and H.~Georgi,
Nucl. Phys. B \textbf{234}, 189-212 (1984).

\bibitem{Shimizu:1984iel}
K.~Shimizu,
Phys. Lett. B \textbf{148}, 418-422 (1984). 

\bibitem{Obukhovsky:1990tx}
I.~T.~Obukhovsky and A.~M.~Kusainov,
Phys. Lett. B \textbf{238}, 142-148 (1990). 

\bibitem{Glozman:1995fu}
L.~Y.~Glozman and D.~O.~Riska,
Phys. Rept. \textbf{268}, 263-303 (1996). 

\bibitem{Glozman:1999vd}
L.~Y.~Glozman,
Nucl. Phys. A \textbf{663}, 103-112 (2000). 

\bibitem{Dai:2003dz}
L.~R.~Dai, Z.~Y.~Zhang, Y.~W.~Yu and P.~Wang,
Nucl. Phys. A \textbf{727}, 321-332 (2003).

\bibitem{Vijande:2004he}
J.~Vijande, F.~Fernandez and A.~Valcarce,
J. Phys. G \textbf{31}, 481 (2005).

\bibitem{Vijande:2009pu}
J.~Vijande and A.~Valcarce,
Phys. Lett. B \textbf{677}, 36-38 (2009).

\bibitem{Valcarce:2005em}
A.~Valcarce, H.~Garcilazo, F.~Fernandez and P.~Gonzalez,
Rept. Prog. Phys. \textbf{68} (2005), 965-1042.

\bibitem{Brambilla:2010cs}
N.~Brambilla, S.~Eidelman, B.~K.~Heltsley, R.~Vogt, G.~T.~Bodwin, E.~Eichten, A.~D.~Frawley, A.~B.~Meyer, R.~E.~Mitchell and V.~Papadimitriou, \textit{et al.}
Eur. Phys. J. C \textbf{71}, 1534 (2011).

\bibitem{Esposito:2014rxa}
A.~Esposito, A.~L.~Guerrieri, F.~Piccinini, A.~Pilloni and A.~D.~Polosa,
Int. J. Mod. Phys. A \textbf{30}, 1530002 (2015).

\bibitem{Chen:2016qju}
H.~X.~Chen, W.~Chen, X.~Liu and S.~L.~Zhu,
Phys. Rept. \textbf{639}, 1-121 (2016).


\bibitem{Chen:2016spr}
H.~X.~Chen, W.~Chen, X.~Liu, Y.~R.~Liu and S.~L.~Zhu,
Rept. Prog. Phys. \textbf{80}, no.7, 076201 (2017).

\bibitem{Esposito:2016noz}
A.~Esposito, A.~Pilloni and A.~D.~Polosa,
Phys. Rept. \textbf{668}, 1-97 (2017).

\bibitem{Guo:2017jvc}
F.~K.~Guo, C.~Hanhart, U.~G.~Mei\ss{}ner, Q.~Wang, Q.~Zhao and B.~S.~Zou,
Rev. Mod. Phys. \textbf{90}, no.1, 015004 (2018).


\bibitem{Olsen:2017bmm}
S.~L.~Olsen, T.~Skwarnicki and D.~Zieminska,
Rev. Mod. Phys. \textbf{90}, no.1, 015003 (2018).


\bibitem{Liu:2019zoy}
Y.~R.~Liu, H.~X.~Chen, W.~Chen, X.~Liu and S.~L.~Zhu,
Prog. Part. Nucl. Phys. \textbf{107}, 237-320 (2019).

\bibitem{Brambilla:2019esw}
N.~Brambilla, S.~Eidelman, C.~Hanhart, A.~Nefediev, C.~P.~Shen, C.~E.~Thomas, A.~Vairo and C.~Z.~Yuan,
Phys. Rept. \textbf{873}, 1-154 (2020).


\bibitem{Chen:2022asf}
H.~X.~Chen, W.~Chen, X.~Liu, Y.~R.~Liu and S.~L.~Zhu,
Rept. Prog. Phys. \textbf{86}, no.2, 026201 (2023). 

\bibitem{LHCb:2021vvq}
R.~Aaij \textit{et al.} [LHCb],
Nature Phys. \textbf{18}, no.7, 751-754 (2022).

\bibitem{LHCb:2021auc}
R.~Aaij \textit{et al.} [LHCb],
Nature Commun. \textbf{13}, no.1, 3351 (2022). 

\bibitem{Pepin:1996id}
S.~Pepin, F.~Stancu, M.~Genovese and J.~M.~Richard,
Phys. Lett. B \textbf{393}, 119-123 (1997). 

\bibitem{Vijande:2003ki}
J.~Vijande, F.~Fernandez, A.~Valcarce and B.~Silvestre-Brac,
Eur. Phys. J. A \textbf{19}, 383 (2004).



\bibitem{Yang:2009zzp}
Y.~Yang, C.~Deng, J.~Ping and T.~Goldman,
Phys. Rev. D \textbf{80}, 114023 (2009).

\bibitem{Deng:2018kly}
C.~Deng, H.~Chen and J.~Ping,
Eur. Phys. J. A \textbf{56}, no.1, 9 (2020).

\bibitem{Yang:2019itm}
G.~Yang, J.~Ping and J.~Segovia,
Phys. Rev. D \textbf{101}, no.1, 014001 (2020).

\bibitem{Tan:2020ldi}
Y.~Tan, W.~Lu and J.~Ping,
Eur. Phys. J. Plus \textbf{135}, no.9, 716 (2020).

\bibitem{Moravcsik_1972}
Michael J Moravcsik, Rep. Prog. Phys. \textbf{35}, 587 (1972).

\bibitem{Meissner:1987ge}
U.~G.~Meissner,
Phys. Rept. \textbf{161}, 213 (1988). 

\bibitem{Machleidt:2017vls}
R.~Machleidt,
Int. J. Mod. Phys. E \textbf{26}, no.11, 1730005 (2017). 

\bibitem{Broniowski:1984zd}
W.~Broniowski and M.~K.~Banerjee,
Phys. Lett. B \textbf{158}, 335 (1985).

\bibitem{Fujiwara:1984pk}
T.~Fujiwara, Y.~Igarashi, A.~Kobayashi, H.~Otsu, T.~Sato and S.~Sawada,
Prog. Theor. Phys. \textbf{74}, 128 (1985).

\bibitem{Igarashi:1985et}
Y.~Igarashi, M.~Johmura, A.~Kobayashi, H.~Otsu, T.~Sato and S.~Sawada,
Nucl. Phys. B \textbf{259}, 721-729 (1985).

\bibitem{Broniowski:1985kj}
W.~Broniowski and M.~K.~Banerjee,
Phys. Rev. D \textbf{34}, 849 (1986).

\bibitem{Meissner:1986vu}
U.~G.~Meissner and I.~Zahed,
Phys. Rev. Lett. \textbf{56}, 1035 (1986).

\bibitem{Birse:1986qc}
M.~C.~Birse,
Phys. Rev. D \textbf{33}, 1934-1950 (1986).




\bibitem{Bando:1987br}
M.~Bando, T.~Kugo and K.~Yamawaki,
Phys. Rept. \textbf{164}, 217-314 (1988).

\bibitem{Harada:2003jx}
M.~Harada and K.~Yamawaki,
Phys. Rept. \textbf{381}, 1-233 (2003).





\bibitem{He:2022su3}
B.~R.~He, M.~Harada and B.~S.~Zou, in preparation.



\bibitem{Hiyama:2003cu}
E.~Hiyama, Y.~Kino and M.~Kamimura,
Prog. Part. Nucl. Phys. \textbf{51}, 223-307 (2003).




\bibitem{ParticleDataGroup:2022pth}
R.~L.~Workman \textit{et al.} [Particle Data Group],
PTEP \textbf{2022}, 083C01 (2022).

\bibitem{Amsler:2018zkm}
C.~Amsler,
Lect. Notes Phys. \textbf{949}, pp.1-277 (2018)
Springer, 2018. 


\end{thebibliography}
\end{document}